\documentclass[a4paper]{article}
\usepackage{amsmath,amssymb, bm, amsthm,amsfonts, colordvi, url, color, mathbbol}
\usepackage{bm, graphicx, graphics, url, natbib, paralist, afterpage, verbatim, textcomp, dsfont}

\bmdefine\ddelta{\delta}
\bmdefine\vvarphi{\varphi}
\bmdefine\ppi{\pi}
\bmdefine\mmu{\mu}
\bmdefine\llambda{\lambda}
\bmdefine\eell{\ell}
\bmdefine\bbeta{\beta}
\bmdefine\ggamma{\gamma}
\bmdefine\vvarepsilon{\varepsilon}
\bmdefine\SSigma{\Sigma}
\bmdefine\LLambda{\Lambda}
\bmdefine\PPsi{\Psi}
\bmdefine\GGamma{\Gamma}
\bmdefine\OOmega{\Omega}

\newtheorem{theorem}{Theorem}

\title{Transfer learning of regression models from a sequence of datasets by penalized estimation}
\author{ {\small \textbf{Wessel N. van Wieringen}$^{1,2,}$\footnote{Corresponding author. Email: w.vanwieringen@amsterdamumc.nl} \,, \textbf{Harald Binder}$^3$}
\\
{\small $^1$ Department of Epidemiology and Data Science, Amsterdam Public Health research institute,}
\\
{\small Amsterdam UMC, location VUmc,  P.O. Box 7057, 1007 MB Amsterdam, The Netherlands}
\\
{\small $^2$ Department of Mathematics, Vrije Universiteit Amsterdam,}
\\
{\small De Boelelaan 1111, 1081 HV Amsterdam, The Netherlands}
\\
{\small $^3$ Institute for Medical Biometry and Statistics,}
\\
{\small Faculty of Medicine and Medical Center, University of Freiburg,}
\\
{\small Stefan-Meier-Str. 26, 79104 Freiburg, Germany}
}
\date{}

\begin{document}
\maketitle

\begin{abstract}
\noindent
Transfer learning refers to the promising idea of initializing model fits based on pre-training on other data. We particularly consider regression modeling settings where parameter estimates from previous data can be used as anchoring points, yet may not be available for all parameters, thus covariance information cannot be reused. A procedure that updates through targeted penalized estimation, which shrinks the estimator towards a nonzero value, is presented. The parameter estimate from the previous data serves as this nonzero value when an update is sought from novel data. This naturally extends to a sequence of data sets with the same response, but potentially only partial overlap in covariates. The iteratively updated regression parameter estimator is shown to be asymptotically unbiased and consistent. The penalty parameter is chosen through constrained cross-validated loglikelihood optimization. The constraint bounds the amount of shrinkage of the updated estimator toward the current one from below. The bound aims to preserve the (updated) estimator's goodness-of-fit on all-but-the-novel data. The proposed approach is compared to other regression modeling procedures. Finally, it is illustrated on an epidemiological study where the data arrive in batches with different covariate-availability and the model is re-fitted with the availability of a novel batch. 
\end{abstract}

\section{Introduction}
In the field of machine learning, many approaches for transfer learning have been developed, which typically amounts to algorithmic approaches for pre-training a model on some other data before moving to the data of interest \citep{Pan2010,Shilo2020}. The idea of transfer learning has also been shown to be more generally useful, e.g., for regression modeling, where it may also be more feasible to get analytical insight into algorithmic approaches \citep{Minami2020}. While transfer learning for regression models allows for differences in the response as well as in the parameters when moving from data set to data set, the framework of transfer learning typically assumes that there is an initial model and that the corresponding structure stays the same when new data become available and need to be incorporated. Transer learning then refers to the updating of knowledge of the model parameter when new information becomes available \citep{Shal2012}. As such it produces a sequence of parameter estimates. The latest estimate takes into account the most recent information but also exploits previously acquired data, while minizing a certain loss function. Examples of tranfer regression techniques abound. Statistical readers will be most familiar with recursive least squares \citep{Plac1950}, mixed models \citep{Verb2009}, or state-space models \citep{Durb2012}. In this context and related, we recently showed how to perform tranfer learning of the normal distribution's precision matrix through targeted ridge penalized estimation \citep{VWie2018}.

In the following, we consider settings where the response stays the same, when moving along a sequence of data sets, but the set of covariates may change. 
For example, data of large epidemiological studies often arrive in batches. Each batch sheds light on the same phenomenon, e.g. environment-disease relation, observed with possibly different covariate information. As such with the arrival of a new batch, the currently learned model of the phenomenon under study needs updating. In this paper we show how updating can be achieved by means of penalized estimation.

Here targeted ridge penalized estimation is proposed as an enrichment of the transfer learning regression toolbox. The presented procedure updates the regression parameter estimate by minimization of the loss function using the novel data and in which the target is formed by the current parameter estimate derived from the hitherto available data. The approach is designed to not depend on covariance estimates of parameters from the previous data, as a change of available covariates will typically invalidate the usefulness of such covariance estimates. Incorporation of multiple competing current estimates, when available, is discussed. Properties, e.g. moments and consistency, of this estimator for tranfer regression learning are proven. All this is then translated to the learning of the logistic regression model. The penalty parameter is chosen via constrained cross-validation to warrant learning and avoid one-off estimation. Transfer regression learning via targeted ridge penalized estimation is compared to standard alternatives. Finally, the potential of the method is illustrated on an epidemiological study.

\section{Methods}
\subsection{Model and penalty structure}
Throughout either a linear or logistic regression model, being the most widely used variants of generalized linear models, are entertained. Both are learned by means of targeted ridge penalized estimation, an extension of ridge regression (inspired by \citealp{Sing1986}). It considers estimating the $p$-dimensional regression coefficient vector $\bbeta$ of the linear regression model, $\mathbf{Y} = \mathbf{X} \bbeta + \vvarepsilon$ with $\vvarepsilon \sim \mathcal{N}(\mathbf{0}_n, \sigma_{\varepsilon}^2 \mathbf{I}_{nn})$, by minimizing the nonzero centered ridge regression loss function:
\begin{eqnarray} \label{form:nonzeroRidgeLoss}
\| \mathbf{Y} - \mathbf{X} \bbeta \|_2^2 + \lambda \| \bbeta - \bbeta_0 \|_2^2,
\end{eqnarray}
with penalty parameter $\lambda$ and nonrandom shrinkage target $\bbeta_0$. An analytic expression for the minimizer of the above loss function exists and gives rise to the nonzero centered ridge regression estimator: $\hat{\bbeta}(\lambda, \bbeta_0) = ( \mathbf{X}^{\top} \mathbf{X} + \lambda \mathbf{I}_{pp})^{-1} ( \mathbf{X}^{\top} \mathbf{Y} + \lambda \bbeta_0)$.  When $\bbeta_0 = \mathbf{0}_{p}$ the traditional ridge regression loss function is recovered. The motivation for this estimator becomes evident after reformulation of the penalized estimation problem above, i.e. the minimization of loss (\ref{form:nonzeroRidgeLoss}), as a constrained estimation problem. When $\bbeta_0 = \mathbf{0}_{p}$, the regression parameter constraint is centered at the origin. But when $\lambda \not= 0$ and $\bbeta_0 \not= \mathbf{0}_{p}$ the constraint on $\bbeta$ is nonzero centered. Of particular interest is the case when an informative choice for $\bbeta_0$ is available. Then, the nonzero centered parameter constraint may contain the true value of $\bbeta$ and, in principle, no shrinkage for the estimate of $\bbeta$ is needed. The targeted ridge estimator then becomes less biased \citep{Sing1986}.

Two alternative interpretations of this estimator are relevant for the presented work. First, the nonzero centered ridge regression can be interpreted as an attempt to correct for the nonzero target. This is easily seen when using the change-of-variable $\ggamma = \bbeta - \bbeta_0$. Substitution of $\ggamma$ in the loss function (\ref{form:nonzeroRidgeLoss}) gives: $\| (\mathbf{Y} - \mathbf{X} \bbeta_0) - \mathbf{X} \ggamma \|_2^2 + \lambda \| \ggamma \|_2^2$. In this $\mathbf{Y} - \mathbf{X} \bbeta_0$ is the residual vector after invoking the prior knowledge. This is regressed on $\mathbf{X}$  to correct for the variance in $\mathbf{Y}$ wrongly attributed to $\mathbf{X}$ by the prior knowledge as condensed in nonzero target $\bbeta_0$. Secondly, from a Bayesian perspective the nonzero centered ridge penalty still corresponds (in the sense that the posterior mean coincides with the optimum of the loss function above) to a normal prior on the elements of the regression parameter $\bbeta$. But this prior now has a mean equal to the corresponding element of $\bbeta_0$.


%

\subsection{Updating the linear regression model} \label{sect:updating_lm}
The targeted ridge regression estimator introduced in the previous section is put to use in the case where multiple data sets come available sequentially at time points $t=1, 2, 3, \ldots$. At each time we fit the linear regression model (now equipped with the index $t$ to separate the data sets: $\mathbf{Y}_t = \mathbf{X}_t \bbeta + \vvarepsilon_t$ with $\vvarepsilon_t \sim \mathcal{N}(\mathbf{0}_{n_t}, \sigma_{\varepsilon}^2 \mathbf{I}_{n_t, n_t})$  using knowledge from the previous case. To this end we use the nonzero centered ridge regression framework. Note that the design matrix $\mathbf{X}_t$ may differ among data sets, in the sense that for the theoretical expos\'{e} the number of covariates is fixed but their settings and the sample size are allowed to change. The effect of these covariates on the response, however, is assumed to be fixed through time.

At the arrival of a new $t$-th data set, the estimates of parameters $\bbeta$ and $\sigma^2$ obtained from the previous $(t-1)$-th data set(s) are available. These may be used as prior knowledge in the updating of the parameter estimates from the $t$-th data set through:
\begin{eqnarray*}
\hat{\bbeta}_t(\lambda_{t}) & = & ( \mathbf{X}_t^{\top} \mathbf{X}_t + \lambda_{t} \mathbf{I}_{pp})^{-1} [ \mathbf{X}_t^{\top} \mathbf{Y}_t + \lambda_{t} \hat{\bbeta}_{t-1}(\lambda_{t-1}, \bbeta_{t-2})],
\end{eqnarray*}
where the previous estimate of $\bbeta$ thus plays the role of $\bbeta_0$. In addition, the $\hat{\bbeta}_{t-1}$ and $\hat{\sigma}^2_{\varepsilon, t-1}$ may be used to choose the penalty parameters by minimization of the MSE plugging in these previous estimates for $\bbeta$ and $\sigma_{\varepsilon}^2$.

The updated ridge linear regression estimator is a linear combination of the observed responses from the subsequent studies. Consequently, it is normally distributed with expectation and variance (see Supplementary Material I for details):
\begin{eqnarray*}
\mathbb{E} \big[ \hat{\bbeta}_t(\lambda_{t}) \big] & = & \sum_{t_h=1}^{t} \Big\{ \prod_{\tau=t_h+1}^t [ \lambda_{\tau} ( \mathbf{X}_\tau^{\top} \mathbf{X}_\tau + \lambda_{\tau} \mathbf{I}_{pp})^{-1}  ]^{I_{\{ t \geq \tau \}}}  \Big\}  \bbeta
\\
& & \,  - \, \sum_{t_h=1}^{t} \Big\{ \prod_{\tau=t_h}^t [ \lambda_{\tau} ( \mathbf{X}_\tau^{\top} \mathbf{X}_\tau + \lambda_{\tau} \mathbf{I}_{pp})^{-1}  ]  \Big\}  \bbeta + \Big\{ \prod_{\tau=1}^t [ \lambda_{\tau}
( \mathbf{X}_\tau^{\top} \mathbf{X}_\tau + \lambda_{\tau} \mathbf{I}_{pp})^{-1}  ]  \Big\}  \bbeta_0,
\\
\mbox{Var} \big[ \hat{\bbeta}_t(\lambda_{t}) \big] & = &  \sigma_{\varepsilon}^2 \sum_{t_h=1}^{t}  \Big[ \prod_{\tau=t_h + 1}^t (\lambda_{\tau}^2 )^{I_{\{ t \geq \tau \}}} \Big] \Big[ \prod_{\tau=t_h}^t ( \mathbf{X}^{\top}_{\tau} \mathbf{X}_{\tau} + \lambda_{\tau} \mathbf{I}_{pp})^{-1} \Big]
\\
& & \qquad \qquad \qquad \qquad \qquad \,  \mathbf{X}^{\top}_{t_h} \mathbf{X}_{t_h} \Big[ \prod_{\tau=t_h}^t ( \mathbf{X}^{\top}_{\tau} \mathbf{X}_{\tau} + \lambda_{\tau} \mathbf{I}_{pp})^{-1} \Big].
\end{eqnarray*}
In the above the $\lambda_{t}$'s have been assumed to be nonrandom.

To provide some intuition assume \textit{i)} every data set is equipped with the same orthonormal design matrix (i.e. $\mathbf{X}_t^{\top} \mathbf{X}_t = \mathbf{I}_{pp}$) and \textit{ii)} $\lambda_{t} = \lambda$ for all $t$. The expectation and variance then reduce to $\mathbb{E} \big[ \hat{\bbeta}_t (\lambda) \big] = \bbeta + \lambda^t ( 1 + \lambda)^{-t} (\bbeta_0  - \bbeta)$ and $\mbox{Var} \big[ \hat{\bbeta}_t(\lambda) \big] = \sigma_{\varepsilon}^2 [ 1 - \lambda^{2t} ( 1 + \lambda)^{-2t} ] \mathbf{I}_{pp}$. In particular, $\lim_{t \rightarrow \infty} \mathbb{E} \big[ \hat{\bbeta}_t (\lambda) \big]  = \bbeta$ and $\lim_{t \rightarrow \infty} \mbox{Var} \big[ \hat{\bbeta}_t (\lambda) \big] = \sigma_{\varepsilon}^2 \mathbf{I}_{pp}$. In words, when the number of novel data sets grows large enough the updated ridge estimator becomes unbiased. Moreover, its variance is smaller than that of the OLS estimator for any $\lambda > 0$. 

More general results on the convergence of the sequence of ridge updated regression estimator and the associated linear predictor can be obtained. To this end notice that the above generated sequence of regression estimators $\{ \hat{\bbeta}_t(\lambda_t, \hat{\bbeta}_{t-1}) \}_{t=1}^{\infty}$ (in which the arguments of $\hat{\bbeta}_{t-1}$ have been dropped to reduce notational clutter) forms a first order Markov chain with continuous state space $\mathbb{R}^p$ as:
\begin{eqnarray*}
\hat{\bbeta}_{t+1} ( \lambda_{t+1}, \hat{\bbeta}_t ) \, | \, \{ \hat{\bbeta}_{t'} (\lambda_t', \hat{\bbeta}_{t-1}) \}_{t'=0}^t & \sim & \hat{\bbeta}_{t+1} ( \lambda_{t+1}, \hat{\bbeta}_t ) \, | \, \hat{\bbeta}_t (\lambda_t, \hat{\bbeta}_{t-1}).
\end{eqnarray*}
Moreover, it is time-homogeneous:
\begin{eqnarray*}
\hat{\bbeta}_{t+\tau+1} ( \lambda_{t+\tau+1}, \hat{\bbeta}_{t+\tau}) \, | \, \hat{\bbeta}_{t+\tau} (\lambda_{t+\tau}, \hat{\bbeta}_{t+\tau-1}), \lambda_{t+\tau} = \lambda  & \sim & \hat{\bbeta}_{t+1} ( \lambda_{t+1}, \hat{\bbeta}_{t} ) \, | \, \hat{\bbeta}_{t} (\lambda_{t}, \hat{\bbeta}_{t-1}), \lambda_{t} = \lambda,
\end{eqnarray*}
for any $\tau \in \mathbb{N}$ and $\lambda_{\tau} > 0$.

The well-known theory of Markov process can now be exploited to show (under some conditions) the asymptotic unbiasedness of the ridge updated linear regression estimator (Theorem \ref{theorem.unbiasedRegressionEstimator}). To this end the following assumption is made:
\begin{compactitem}
\item[\underline{\textit{A1.linear}}] There exists an infinite sequence of studies into the linear relationship between a continuous response and a set of covariates. The data from these studies, $\{ \mathbf{X}_t, \mathbf{Y}_t \}_{t=1}^{\infty}$, are used to fit the linear regression model by means of the updated ridge linear regression estimator, which yields the sequence of estimators $\{ \hat{\bbeta}_{t} (\lambda_t, \hat{\bbeta}_{t-1}) \}_{t=1}^{\infty}$ which is initiated by an arbitrary, nonrandom $\bbeta_0$.
\end{compactitem}
The theorem's proof, deferred -- as all proofs -- to the Supplementary Material II, requires to show that the updating process has a stationary distribution with expectation equal to $\bbeta$.

\begin{theorem} (Asymptotic unbiasedness of updated ridge linear regression estimator) \label{theorem.unbiasedRegressionEstimator}
\\
Adopt assumption \textit{A1.linear}. Let $T \in \mathbb{N}$ be sufficiently large. Then, $\lim_{t \rightarrow \infty} \mathbb{E} [\hat{\bbeta}_{t+1} ( \lambda_{t+1}, \hat{\bbeta}_{t} )]= \bbeta + \mathbf{u}$ for some $\mathbf{u} \in \cap_{t=T}^{\infty} \mbox{null}(\mathbf{X}_t)$, where $\mbox{null}(\mathbf{X}_t)$ denotes the null space of the linear map induced by $\mathbf{X}_t$. If $\cap_{t=T}^{\infty} \mbox{null}(\mathbf{X}_t) = \emptyset$, then $\mathbf{u} = \mathbf{0}_p$ and, consequently, the updated ridge linear regression estimator is asymptotically unbiased.
\end{theorem}

\noindent
Asymptotic unbiasedness can also be shown for the linear predictor (Theorem \ref{theorem.unbiasedRegressionPredictor}), which the restriction on the intersection on null spaces of the design matrices is not needed.

\begin{theorem} (Asymptotic unbiasedness of updated ridge linear regression predictor) \label{theorem.unbiasedRegressionPredictor}
\\
Adopt assumption \textit{A1.linear}. Let $\mathbf{X}_{\mbox{{\tiny new}}}$ be the design matrix with covariate information on novel samples for which a prediction is needed. The updated linear predictor associated with the updated ridge linear regression estimator is then asymptotically unbiased: $\lim_{t \rightarrow \infty} \mathbb{E} [\mathbf{X}_{\mbox{{\tiny new}}} \hat{\bbeta}_{t+1} ( \lambda_{t+1}, \hat{\bbeta}_{t} )] = \mathbf{X}_{\mbox{{\tiny new}}} \bbeta$.
\end{theorem}

\noindent
A few notes on these theorems are to be pointed out. First, both theorems describe asymptotic behavior. But they do so not in the traditional sense, in which the sample size goes by one at the time. Here that may be if a new study with a degenerated sample size is included. But as most studies are expected to have a sample size exceeding one, the limit is reached in larger step sizes. Moreover, the data of the new study is not added to the data of the previous study, rather the former is weighed against a summary obtained from the latter: the current updated ridge linear regression estimator.

Another, more important note concerns the role of the penalty parameters $\{ \lambda_t \}_{t=1}^{\infty}$ in  Theorems \ref{theorem.unbiasedRegressionEstimator} and \ref{theorem.unbiasedRegressionPredictor}. The asymptotic unbiasedness of estimator and predictor is independent of the choice of the penalty parameters. Practically, this means that whether one chooses the penalty by means of cross-validation or some other means the asymptotic unbiasedness of the updated ridge linear regression estimator and predictor is warranted. Moreover, irrespectively of the amount of penalization applied at each update Theorems  \ref{theorem.unbiasedRegressionEstimator} and \ref{theorem.unbiasedRegressionPredictor} hold.

In addition to asymptotic unbiasedness the updated ridge linear regression estimator can be shown to be consistent (Theorem \ref{theorem:consistency}), although under assumptions on the penalty parameter sequence.

\begin{theorem} \label{theorem:consistency} \mbox{(Consistency of the updated ridge linear regression estimator and predictor)}
\\
Adopt assumption \textit{A1.linear}. Let $T \in \mathbb{N}$ be sufficiently large and $\cap_{t=T}^{\infty} \mbox{null}(\mathbf{X}_t) = \emptyset$. Assume for all $t \geq T$ that $\lambda_{t}^{-1} \gg \lambda_{t}^{-2}$ and $\lambda_{t} > 2 [d_1 (\mathbf{X}_t)]^{2}$ with $d_1 (\mathbf{X}_t)$ the largest singular value of $\mathbf{X}_t$. Then, the updated ridge linear regression estimator and the associated linear predictor  are consistent, i.e. $p$-$\lim_{t \rightarrow \infty} \hat{\bbeta}_{t+1}(\lambda_{t+1}, \hat{\bbeta}_t) = \bbeta$ and $p$-$\lim_{t \rightarrow \infty} \mathbf{X}_{\mbox{{\tiny new}}} \hat{\bbeta}_{t+1}(\lambda_{t+1}, \hat{\bbeta}_t)  = \mathbf{X}_{\mbox{{\tiny new}}} \bbeta$, respectively.
\end{theorem}

The proposed updating scheme may be seen as a frequentist analogue of Bayesian updating \citep{Berg2013}. In the Bayesian narrative, the ridge penalty corresponds to a normal prior. The resulting posterior of $\bbeta$ is also a normal distribution. When updating, this normal posterior then serves as (normal) prior for the next update of $\bbeta$. The prior for the $t+1$-th update then becomes $\bbeta_{t+1} \, | \, \sigma^2 \sim \mathcal{N}[  \bbeta_{t}(\lambda_{t}, \bbeta_{t-1}), \sigma^2 \lambda_{t+1}^{-1} \mathbf{I}_{pp}]$, which yields a posterior mean $\mathbb{E}[ \bbeta_{t+1} \, | \, \mathbf{Y}_{t+1}, \mathbf{X}_{t+1}, \sigma^2, \bbeta_{t}(\lambda_{t}, \beta_{t-1})]$  coinciding with the frequentist estimator $\hat{\bbeta}_{t+1}(\lambda_{t+1}, \bbeta_{t})$.

Finally, the target for the estimation of the regression parameter from the current data need not be the most recently updated ridge regression estimator. It may be replaced by alternative estimators obtained from the preceding data sets. These targets may even be simultaneously available prior to the estimation of the regression parameter from the current data. One would then preferably have the current data choose among these targets. This can be done straightforwardly within the current framework. Hereto replace the targeted ridge penalty by what could be considered a mixture of such penalties. For $G$, $G \in \mathbb{N}$, such targets $\bbeta_{0,t,1}, \ldots, \bbeta_{0,t,G}$, this yields the loss function: $\| \mathbf{Y}_t - \mathbf{X}_t \bbeta \|_2^2 + \lambda_t \sum\nolimits_{g=1}^G \alpha_{t,g} \| \bbeta - \bbeta_{0,t,g} \|_2^2$, where all $\alpha_{t,g} \geq 0$ and $\sum_{g=1}^G \alpha_{t,g} = 1$. This loss is minimized with respect to $\bbeta$ by:
\begin{eqnarray*}
\hat{\bbeta} (\lambda_t, \alpha_{t,1}, \ldots, \alpha_{t,G}) & = & (\mathbf{X}^{\top} \mathbf{X} + \lambda_t \mathbf{I}_{pp})^{-1} (\mathbf{X}^{\top} \mathbf{Y} + \lambda_t \sum\nolimits_{g=1}^G \alpha_{t,g} \bbeta_{0,t,g} ).
\end{eqnarray*}
The targets are effectively averaged weightedly to form a novel target. The weight of this average are unknown. Like the penalty parameter $\lambda_t$ the weights are tuning parameters and are to be chosen prior to the estimation, e.g. through a procedure like cross-validation (cf. Section \ref{sect.CV}). 

\subsection{Updating the logistic regression model}
The results of the previous section carry over to generalized linear models. This is exemplified here for the logistic regression model. Hereto the same experimental set-up as for the linear regression model is assumed to apply, now with a binary response: $Y_i \in \{ 0, 1\}$ for $i=1, \ldots, n$. The probability of a `1' -- and thereby that of a `0' -- is modeled by the logistic link function: $P(Y_i = 1)  = \exp ( \mathbf{X}_{i,\ast} \bbeta) / [1 + \exp ( \mathbf{X}_{i,\ast} \bbeta) ]$ for every $i$. Again the regression parameter $\bbeta$ is estimated by maximization of the loglikelihood augmented with a nonzero centered ridge penalty:
\begin{eqnarray*}
\hat{\bbeta}(\lambda) & = & \arg \max_{\bbeta \in \mathbb{R}^p} \sum\nolimits_{i=1}^n  \big\{ Y_i \mathbf{X}_{i,\ast} \bbeta - \log [ 1 + \exp(\mathbf{X}_{i, \ast} \bbeta) ] \big\} - \lambda \| \bbeta - \bbeta_0 \|_2^2.
\end{eqnarray*}
The corresponding estimating equation is:
\begin{eqnarray} \label{form:logisticEstimatingEquation}
\mathbf{X}^{\top} [\mathbf{Y} - \vec{\mathbf{g}}^{-1}( \mathbf{X}; \bbeta)] - \lambda (\bbeta - \bbeta_0) & = & \mathbf{0}_p,
\end{eqnarray}
where $\vec{\mathbf{g}}^{-1}( \mathbf{X}; \bbeta)  = [g^{-1}( \mathbf{X}_{1, \ast}; \bbeta), \ldots, g^{-1}( \mathbf{X}_{n, \ast}; \bbeta)]^{\top}$ with link function $g(\cdot; \cdot)$ defined by $g^{-1}(\cdot; \cdot) = \exp(\cdot; \cdot) / [1 + \exp(\cdot; \cdot)]$. The  root of Equation (\ref{form:logisticEstimatingEquation}) is found by means of a Newton-Raphson algorithm, reformulated as an iteratively re-weighted least squares algorithm. This requires a minor modification from that presented in \cite{Scha1984,LeCe1992} for the zero-centered ridge logistic regression estimator. Starting from an initial guess $\hat{\bbeta}^{(0)}$ the estimate is updated by:
\begin{eqnarray*}
\hat{\bbeta}^{(k+1)}(\lambda) & = & (\mathbf{X}^{\top} \mathbf{W}  \mathbf{X} +  \lambda \mathbf{I}_{pp} )^{-1} ( \mathbf{X}^{\top} \mathbf{W}  \mathbf{Z} + \lambda \bbeta_0),
\end{eqnarray*}
where $\mathbf{W}$ diagonal with $(\mathbf{W})_{ii} = \exp(\mathbf{X}_i \hat{\bbeta}^{(k)}  ) [ 1 + \exp(\mathbf{X}_i \hat{\bbeta}^{(k)} ) ]^{-2}$ and the adjusted response variable $\mathbf{Z} = \{ \mathbf{X} \hat{\bbeta}^{(k)} +  \mathbf{W}^{-1} [\mathbf{Y} - \vec{\mathbf{g}}^{-1}( \mathbf{X}; \bbeta^{(k)})] \}$. This recursive formula is applied until convergence, which yields the desired nonzero centered ridge logistic regression estimate.

Now assume data $\{\mathbf{X}_{t}, \mathbf{Y}_t \}_{t=1}^{\infty}$ from a sequence of studies into the same logistic regression model, i.e. with a common  regression parameter $\bbeta$, but possibly different design matrices are available. From each study the parameter is estimated by means of nonzero centered ridge logistic regression estimator with the estimate of the previous study as shrinkage target. The resulting estimate  $\hat{\bbeta}_{t+1}[\lambda_{t+1}, \hat{\bbeta}_{t}(\lambda_t, \hat{\bbeta}_{t-1})]$ is that $\bbeta$ which solves:
\begin{eqnarray*}
\mathbf{X}^{\top}_{t+1} [ \mathbf{Y}_{t+1} - \vec{\mathbf{g}}^{-1}( \mathbf{X}_{t+1}; \bbeta) ] - \lambda_{t+1} [\bbeta - \hat{\bbeta}_{t}(\lambda_t, \hat{\bbeta}_{t-1})] & = & \mathbf{0}_p.
\end{eqnarray*}
This yields a sequence of estimators $\{\hat{\bbeta}_{t}(\lambda_t, \hat{\bbeta}_{t-1}) \}_{t=2}^{\infty}$, in which the (again) the arguments of the $\hat{\bbeta}_{t-1}$ have been dropped to reduce notational clutter. The $t+1$-th estimator $\hat{\bbeta}_{t+1} (\lambda_{t+1}, \hat{\bbeta}_{t})$ of this sequence relates, after convergence of the iteratively re-weighted least squares algorithm, to its predecessor by:
\begin{eqnarray*}
\hat{\bbeta}_{t+1}(\lambda_{t+1}, \hat{\bbeta}_{t} ) & = & (\mathbf{X}_{t+1}^{\top} \mathbf{W}_{t+1}  \mathbf{X}_{t+1} +  \lambda_{t+1} \mathbf{I}_{pp} )^{-1} [ \mathbf{X}_{t+1}^{\top} \mathbf{W}_{t+1}  \mathbf{Z}_{t+1} + \lambda_{t+1} \hat{\bbeta}_{t}(\lambda_t, \hat{\bbeta}_{t-1})],
\end{eqnarray*}
with $\mathbf{W}_{t+1}$ and $\mathbf{Z}_{t+1}$ now involving $\mathbf{X}_{t+1}$ and $\mathbf{Y}_{t+1}$.

The thus defined sequence of updated logistic regression estimators $\{\hat{\bbeta}_{t}(\lambda_t, \hat{\bbeta}_{t-1} ) \}_{t=2}^{\infty}$ can again be seen as being generated by a $1^{\mbox{{\tiny st}}}$ order Markov chain with $\mathbb{R}^p$ as its state space. Like before this fact is exploited to show asymptotic properties of the updated ridge logistic regression estimator. Throughout the presented asymptotic results the following assumption is adopted:
\begin{compactitem}
\item[\underline{\textit{A2.logistic}}] Assume an infinite sequence of studies into the generalized linear relationship between a binary response and a set of covariates. The data from these studies, $\{ \mathbf{X}_t, \mathbf{Y}_t \}_{t=1}^{\infty}$, are used to fit the logistic regression model by means of the updated ridge logistic regression estimator, which yields the sequence of estimators $\{ \hat{\bbeta}_{t} (\lambda_t, \hat{\bbeta}_{t-1}) \}_{t=1}^{\infty}$ which is initiated by an arbitrary, nonrandom $\bbeta_0$.
\end{compactitem}
Theorems \ref{theorem.unbiasedLogisticRegressionEstimator} and \ref{theorem.unbiasedLogisticRegressionPredictor} below state the asymptotic unbiasedness of the updated estimator and predictor, respectively, while Theorem \ref{theorem:consistencyLogistic} states their consistency.

\begin{theorem} (Asymptotic unbiasedness of updated logistic regression estimator) \label{theorem.unbiasedLogisticRegressionEstimator}
\\
Adopt assumption \textit{A2.logistic}. Let $T \in \mathbb{N}$ be sufficiently large. Then, $\lim_{t \rightarrow \infty} \mathbb{E} [\hat{\bbeta}_{t+1} ( \lambda_{t+1}, \hat{\bbeta}_{t} )]= \bbeta + \mathbf{u}$ for some $\mathbf{u} \in \cap_{t=T}^{\infty} \mbox{null}(\mathbf{X}_t)$. If $\cap_{t=T}^{\infty} \mbox{null}(\mathbf{X}_t) = \emptyset$, then $\mathbf{u} = \mathbf{0}_p$ and, consequently, the updated ridge logistic regression estimator is asymptotically unbiased.
\end{theorem}

\begin{theorem} (Asymptotic unbiasedness of updated logistic regression predictor) \label{theorem.unbiasedLogisticRegressionPredictor}
\\
Adopt assumption \textit{A2.logistic}. Let $\mathbf{X}_{\mbox{{\tiny new}}}$ be the design matrix with covariate information on novel samples for which a prediction is needed. The updated linear predictor formed from the updated ridge logistic regression estimator is then asymptotically unbiased: $\lim_{t \rightarrow \infty} \mathbb{E} [\mathbf{X}_{\mbox{{\tiny new}}} \hat{\bbeta}_{t+1} ( \lambda_{t+1}, \hat{\bbeta}_{t} )] = \mathbf{X}_{\mbox{{\tiny new}}} \bbeta$.
\end{theorem}

\begin{theorem} \label{theorem:consistencyLogistic} \mbox{(Consistency of the updated ridge logistic regression estimator and predictor)}
\\
Adopt assumption \textit{A2.logistic}. Let $T \in \mathbb{N}$ be sufficiently large and $\cap_{t=T}^{\infty} \mbox{null}(\mathbf{X}_t) = \emptyset$. Assume for all $t \geq T$ that $\lambda_{t}^{-1} \gg \lambda_{t}^{-2}$ and $\lambda_{t} > 2 [d_1 (\mathbf{X}_t)]^{2}$ with $d_1 (\mathbf{X}_t)$ the largest singular value of $\mathbf{X}_t$. Then, the updated ridge logistic regression estimator and the associated linear predictor are consistent, i.e. $p$-$\lim_{t \rightarrow \infty} \hat{\bbeta}_{t+1}(\lambda_{t+1}, \hat{\bbeta}_t) = \bbeta$ and $p$-$\lim_{t \rightarrow \infty} \mathbf{X}_{\mbox{{\tiny new}}} \hat{\bbeta}_{t+1}(\lambda_{t+1}, \hat{\bbeta}_t) = \mathbf{X}_{\mbox{{\tiny new}}} \bbeta$, respectively.
\end{theorem}

The updated ridge logistic regression estimator too is a frequentist analogue of a Bayesian updating scheme. Now the analogy is not exact but approximate. This is due to the fact that the normal prior is not conjugate for the logistic likelihood and, consequently, the posterior is not a well-known and characterized distribution. Nonetheless, the latter may be approximated in a Laplacian manner by a normal distribution (cf. \cite{Bish2006}), for which the Bernstein-Von Mises theorem \cite{VdVa1998} provides conditions for its quality. The mean of this approximating normal corresponds to the mode of posterior (i.e. the updated ridge logistic regression estimator), while its variance relates to the curvature of the posterior at its mode. This normal-like posterior then serves as prior when updating the knowledge on the logistic regression parameter at the arrival of a novel data set.

The targeted ridge logistic regression estimator may also be modified to accommodate multiple targets acquired from different estimators. Again, this amounts to averaging the targets weightedly.

\subsection{Choice of the tuning parameter} \label{sect.CV}
Informative choices for the tuning parameters, $\{ \lambda_t \}_{t=1}^{\infty}$ and $\bbeta_0$, of updated ridge estimators are presented. The ridge penalty parameter of a novel data set is chosen through a form of cross-validation. Cross-validation chooses the penalty parameter that yields the best performance of the model on novel data. In absence of novel data part of data, referred to as the \textit{test} or \textit{left-out} data, is set aside to serve as such. The remaining data, called the \textit{training} or \textit{left-in} data, are used to learn the model that is evaluated on the test data. The thus acquired performance depends on the selection of the test data, and may accidently yield an overly optimistic performance. To remove the effect of the particular the data are split into training and test data repetitively, say, $K$ times, thus giving rise to $K$-fold cross-validation. Splitting may be done randomly but is done here in a stratified manner. The latter procedure ensures approximate equally sized test data sets and the single occurrence of each sample in a test data set. The performance averaged over the test data sets is considered to be representative for novel data. The chosen penalty parameter optimizes this performance. The model's performance is usually measured by its fit  on the test data. For the linear regression model this amounts to the cross-validated sum-of-squared ${K}^{-1} \sum_{k=1}^K \| \mathbf{Y}_{t}^{(k)} - \mathbf{X}_{t}^{(k)} \hat{\bbeta}_{t}^{(-k)} (\lambda_t) \|_2^2$ where $\mathbf{Y}_{t}^{(k)}$ and $\mathbf{X}_{t}^{(k)}$ are formed by subsetting $\mathbf{Y}_{t}$ and $\mathbf{X}_{t}$ to the $k$-th test data, while $\hat{\bbeta}_{t}^{(-k)}$ is the ridge regression estimator obtained from all but the $k$-th test data. In case of logistic regression the cross-validated likelihood ${K}^{-1} \sum_{k=1}^K \mathcal{L} [ \mathbf{Y}_{t}^{(k)}, \mathbf{X}_{t}^{(k)},  \hat{\bbeta}_{t}^{(-k)} (\lambda_t) ]$  is used as a performance of measure. The optimal $\lambda_t$ is found by minimization/maximization of the appropriate criterion using standard machinery.

The proposed cross-validation procedure does not take into account the data sets prior to the $t$-th one. Unforeseen circumstances may have lead to a realization of the $t$-th data set that is not representative. For instance, the response-covariate relation may be absent, i.e. $\bbeta = \mathbf{0}_p$, in this particular data set. Cross-validation, which optimizes the predictive performance in the $t$-th data set, is likely to select a rather small $\lambda_t$ that will not shrink towards the updated ridge regression estimator obtained from the previous $t-1$ data sets. As such the information from these prior data sets accumulated in the updated ridge regression estimator is ignored. With the $t$-th data set being the odd-one-out, estimation of the regression parameter from the $t+1$-th data set starts anew with an uninformative shrinkage target. To safeguard against such phenomena the cross-validated performance measure is still optimized with respect to the penalty parameter but with a constraint on the detoriation of the fit on the preceding data sets. The minimization takes place over the set $\mathcal{D}$ defined as:
\begin{eqnarray*}
\Big\{ \lambda_t > 0 : (1 - f_t)  K^{-1} \sum\nolimits_{k=1}^K \sum\nolimits_{\tau=1}^{t-1} \| \mathbf{Y}_{\tau} - \mathbf{X}_{\tau} \hat{\bbeta}_{t}^{(-k)} (\lambda_{t}) \|_2^2  \, \leq \, \sum\nolimits_{\tau=1}^{t-1} \| \mathbf{Y}_{\tau} - \mathbf{X}_{\tau} \hat{\bbeta}_{t-1} (\lambda_{t-1}) \|_2^2 \Big\}
\end{eqnarray*}
with $f_t = n_t / (\sum_{\tau=1}^{t} n_{\tau})$ denoting the sample size fraction of the $t$-th data set in relation to the sample size accumulated over all data sets up to $t$. In the optimization the constraint is enforced using a barrier function. Constraining the choice of $\lambda_t$ to $\mathcal{D}$ ensures that the fit of previously acquired data set does not detoriate more than the fraction $f_t$. In the presence of multiple targets the constraint simply demarcates a viable subspace of the penalty parameters $\lambda_t$ and the $\alpha_{t,g}$'s. Finally, a similar constraint can be conceived for the choice of the ridge logistic regression estimator's penalty parameter through cross-validation. The sum-of-squares are then to be replaced by the (minus) loglikelihoods.

An informative choice on $\bbeta_0$ that initiates the updating may be obtained from profound knowledge of the relationship described by the linear regression model. Usually, however, such knowledge is at best present in tacit form. One way out could be literature providing univariate estimates of (some of) the $\beta_j$'s that comprise $\bbeta$. An alternative would `sacrifice' the first data set to obtain an initial estimate with which the updating is initiated. Traditional ridge regression (with a zero target) or, following literature, univariate estimates from covariate-wise simple linear regression could be used to arrive at this initial estimate. Due to the shrinkage to zero the former tends to underestimate the elements of $\bbeta$, whereas the latter has the opposite result (as confounding covariates are omitted). Alternatively, an estimate from other transfer regression learning procedures -- preferably an unbiased one -- can be used to initiate the updating.

Initiation is required not only at the initial time point. With the onset of the big data age, more and more information is registered over time. Consequently, `early' data sets comprise a limited number of covariates, while more recent ones comprise many additional ones. Or, it may be too costly to measure certain covariates at each instance but the budget allows for their measurement at (say)  every other study. Hence, in practice not every covariate will be present in each data set. A pragmatic choice of the target is than to use the latest estimates element-wise.

\section{Comparisons}
Two comparisons are conducted. The first illustrates that updating is beneficial compared to \textit{de novo} estimation. Secondly, updating is  contrasted, both \textit{in papyro} and \textit{in silico}, to estimation from the pooled data

\subsection{Regular ridge regression} \label{sect:comparison2regularRidge}
In a small simulation study the behavior of the updated ridge linear regression estimator is contrasted to its traditional ridge counterpart. First, an initial data set is generated from $\bbeta$ is estimated with the traditional ridge regression estimator in combination with a LOOCV chosen penalty parameter. This estimate serves as the $\bbeta_0$ to initiate the updated ridge regression. Now sequentially another 25 data sets are drawn. From each $\bbeta$ is estimated using both the traditional and updated ridge estimators with LOOCV (without any constraint but positivity) for penalty parameter determination. For the latter the previous updated ridge regression estimate is used as target. This process is repeated a hundred times. 

All data sets are drawn as follows. Throughout the $j$-th element of the vector of regression coefficients $\bbeta$ is given by $\beta_j = (j-50)/20$ for $j=0, 1, 2, \ldots, 100$. Each element of the $(n \times p)$-dimensional $t$-th design matrix $\mathbf{X}_t$ with $n=25$ is drawn from the standard normal distribution. Then, $\mathbf{Y}_t = \mathbf{X}_t \bbeta + \vvarepsilon_t$ with each element of $\vvarepsilon_t$ sampled from $\mathcal{N}(0, 0.04)$. Hence, $\mathbf{X}_t$, $\vvarepsilon_t$, and consequently $\mathbf{Y}_t$, are generated anew at each $t$. 

\begin{figure*}[b!]
\centering
\begin{tabular}{ll}
\includegraphics[angle=0, scale=0.41]{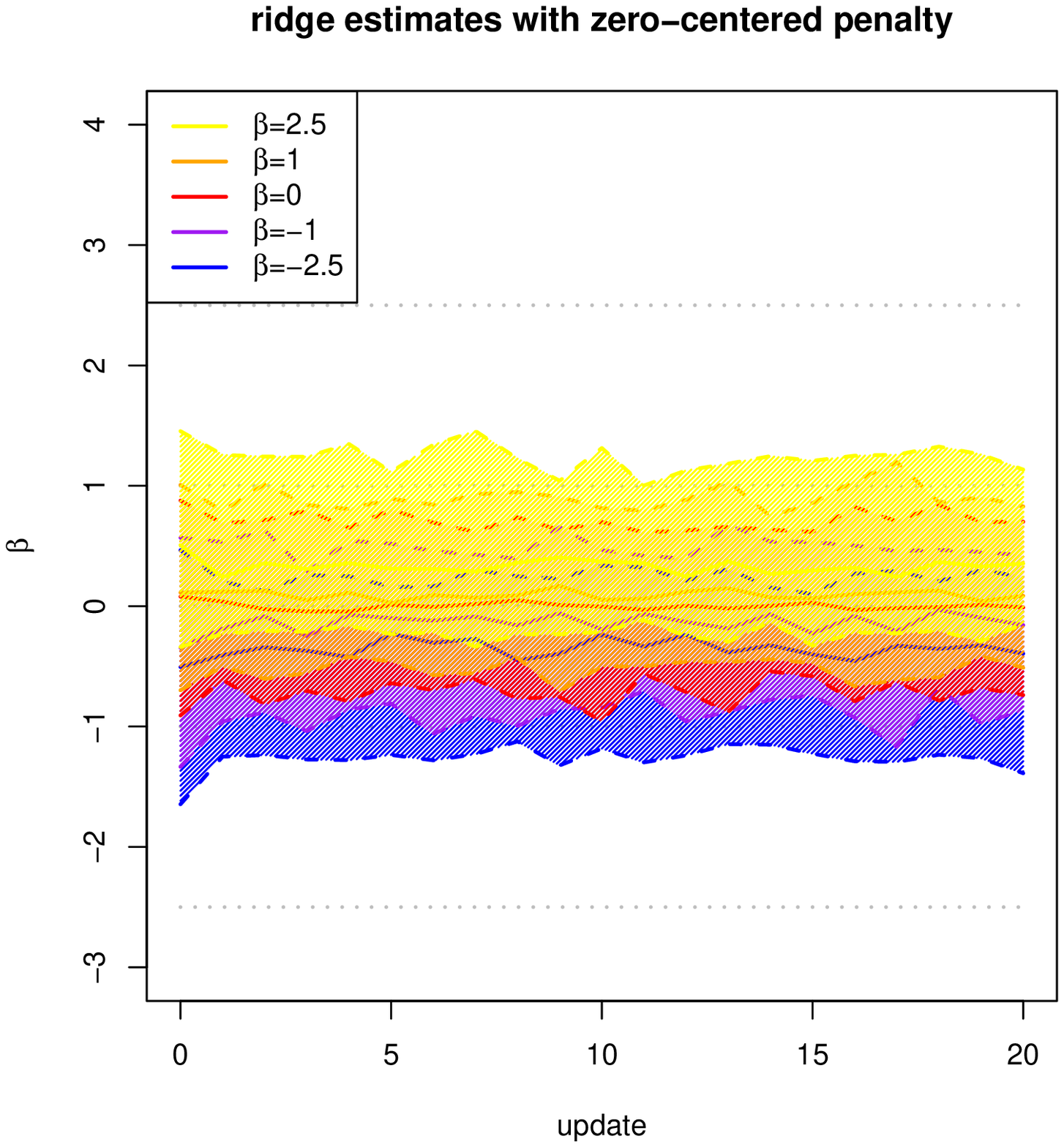}
&
\includegraphics[angle=0, scale=0.41]{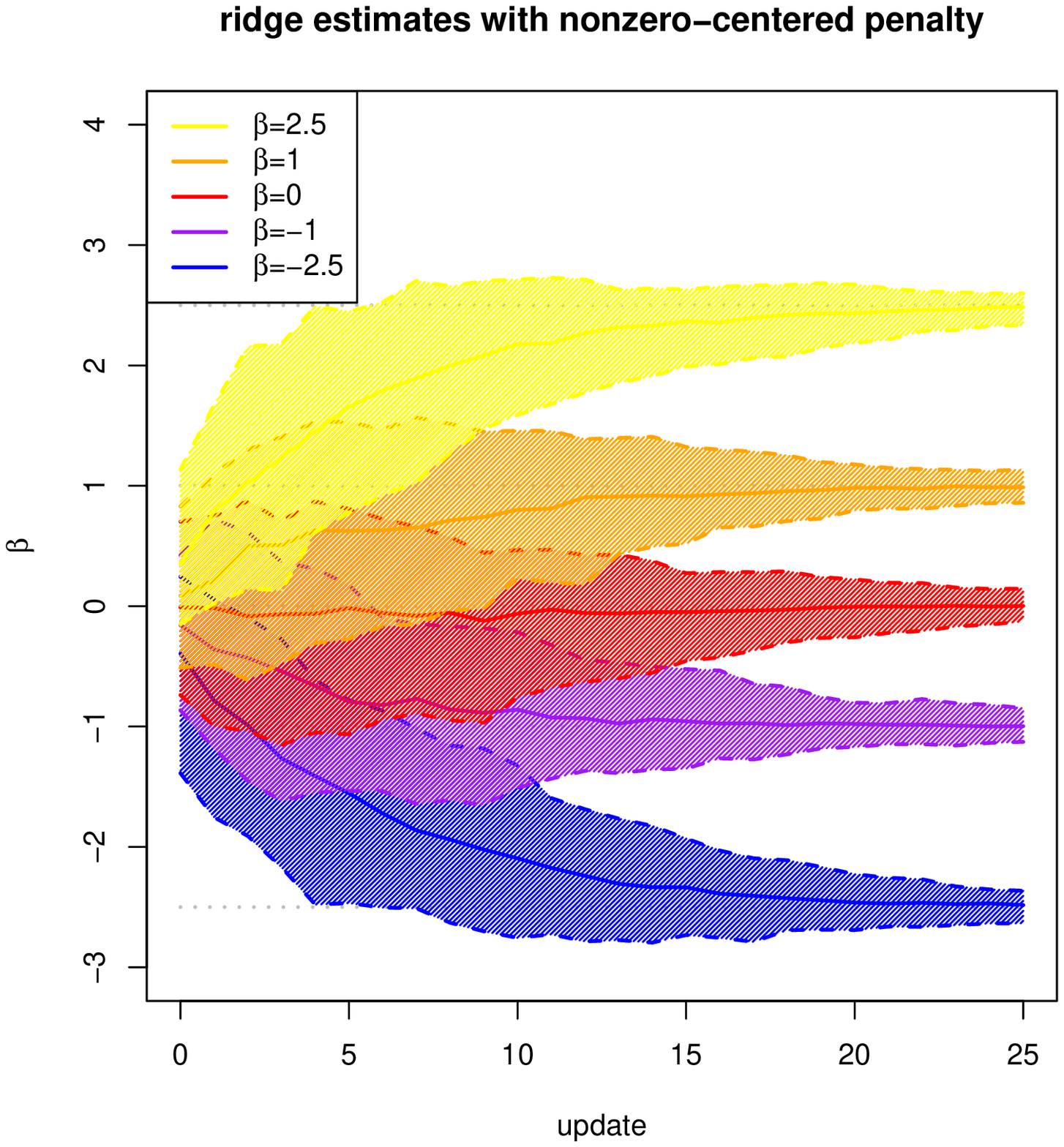}
\\
\includegraphics[angle=0, scale=0.41]{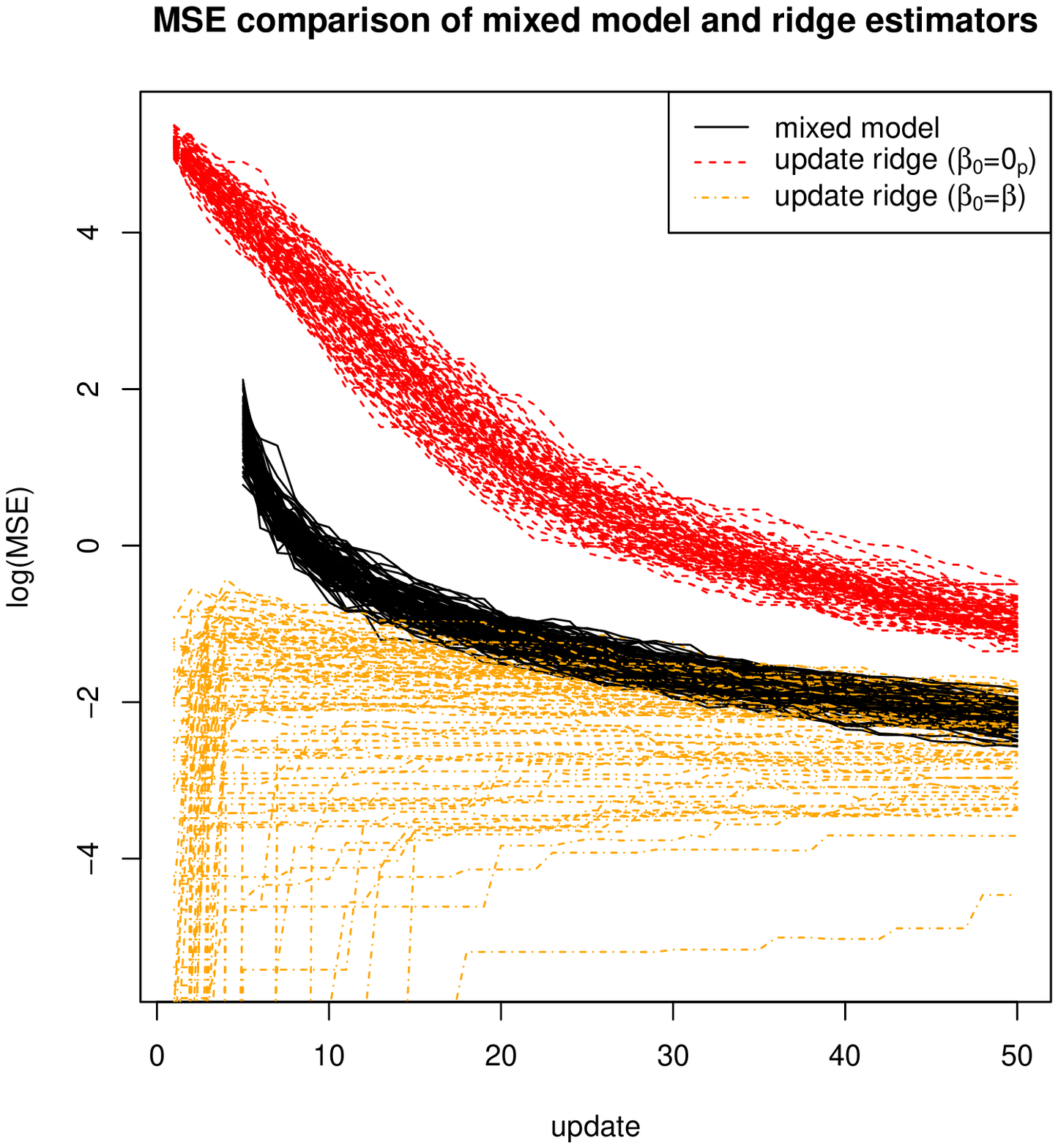}
&
\includegraphics[angle=0, scale=0.41]{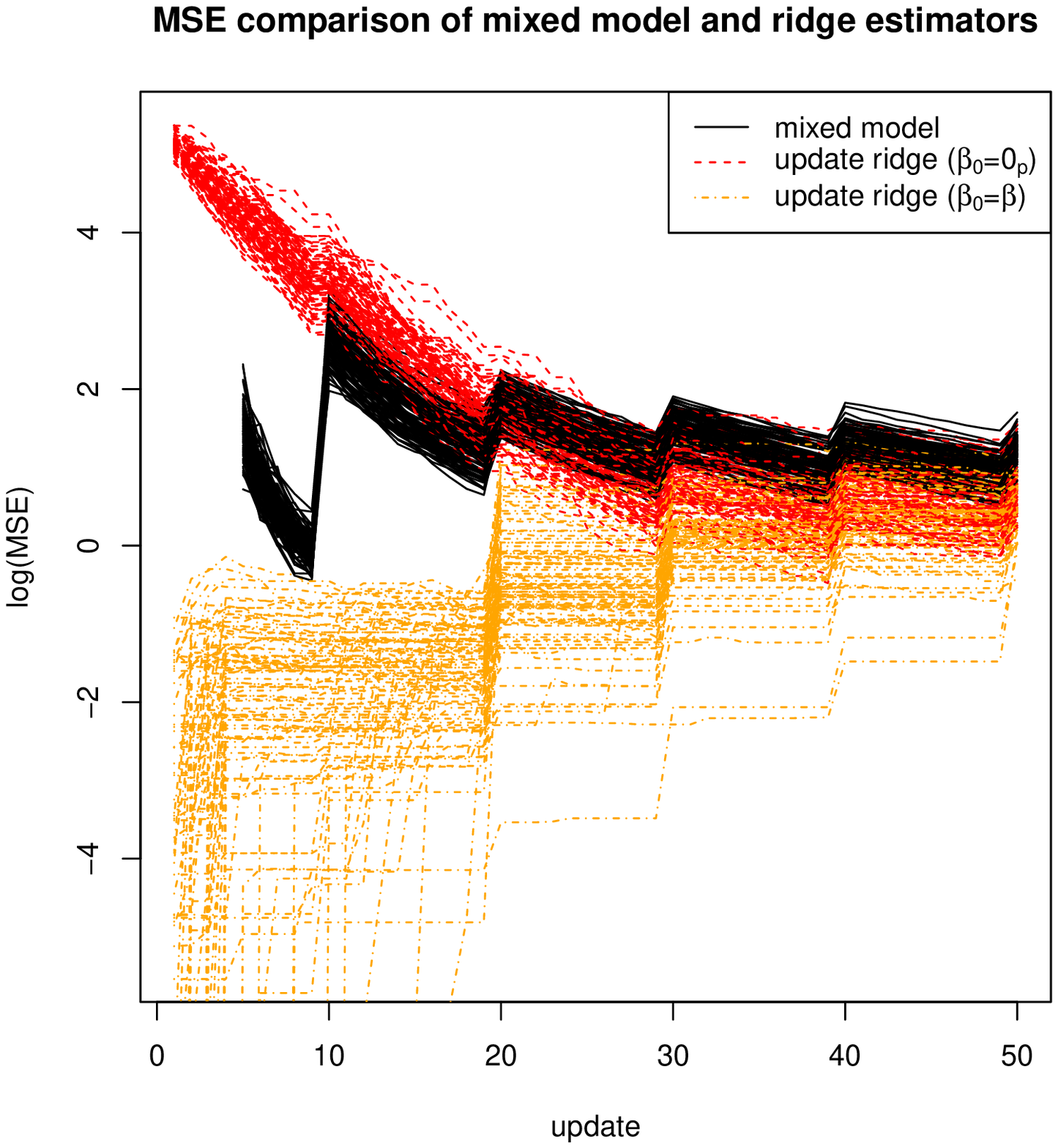}
\end{tabular}
\caption{The top panels show the $(5\%,  95\%$-quantile intervals of the  traditional (left) and updated (right) ridge estimates of $\beta_{j}$ with $j \in \{ 0, 21, 51, 71, 101 \}$ plotted against $t$. The solid, colored line inside these intervals is the corresponding $50\%$ quantile. The dotted, grey lines are the true values of the $\beta_j$'s. The bottom panels show the mean squared errors over hundred runs of the mixed model's maximum likelihood and updated ridge regression estimators (initiated with $\bbeta_0 = \mathbf{0}_p$ and $\bbeta_0 = \bbeta$)
plotted against $t$. All data sets are sampled from the same regression model, but in the right  panel every tenth data set is sampled from an empty model.} \label{fig.regular2updateRidge}
\hfill{}
\end{figure*}

In Figure \ref{fig.regular2updateRidge} the $5\%$, $50\%$ and $95\%$ quantiles of the traditional and updated ridge estimates of $\beta_{j}$ with $j \in \{ 0, 21, 51, 71, 101 \}$ are plotted against $t$. These quantiles of the traditional ridge estimates of these elements of $\bbeta$ are constant over $t$ (left panel of Figure \ref{fig.regular2updateRidge}). Those of the update ridge estimates (right panel of Figure \ref{fig.regular2updateRidge}) clearly improve as $t$ increases. the improvement is two-fold: \textit{i)} they become less biased, and \textit{ii)} the distance between the $5\%$ and $95\%$ quantiles vanishes.

\subsection{Mixed model} \label{sect:comp2mixedModel}
A natural and widespread alternative to the proposed approach would be to employ a mixed model that includes a data set-specific random effect of the covariates. The mixed model is then refitted at the arrival of a new data set, each time to an enlarged data set encompassing the newly arrived data and all those that preceded it. This may be computationally demanding but is considered a mere practical nuisance for the moment. To make matters more precise consider for the data set formed from all those up to the $t$-th one the mixed model $\mathbb{Y}_t = \mathbb{X}_t \bbeta + \mathbb{Z}_t \mathbb{G}_t + \mbox{{\Large $\mathbb{e}$}}_t$ with:
\begin{compactitem}
\item the vector $\mathbb{Y}_t = (\mathbf{Y}_1^{\top}, \ldots, \mathbf{Y}_t^{\top})^{\top}$ of length  $\tilde{n}_t$-dimension where $\tilde{n}_t = n_1 + \ldots + n_t$,
\item the $(\tilde{n}_t \times p)$-dimensional matrix $\mathbb{X}_t = (\mathbf{X}_1^{\top}, \ldots, \mathbf{X}_t^{\top})^{\top}$,
\item the $(\tilde{n}_t \times tp)$-dimensional matrix $\mathbb{Z}_t$ constructed as a $t \times t$ block matrix with $\mbox{diag}(\mathbb{Z}_t) = (\mathbf{X}_1, \ldots, \mathbf{X}_t)$ and off-diagonal blocks all equalling the zero matrix of appropriate dimensions,
\item the vector $\mathbb{G}_t = (\ggamma_1^{\top}, \ldots,\ggamma_t^{\top} )^{\top}$ of length $t p$ with the data set-specific random effects, and
\item the error vector $\mbox{{\Large $\mathbb{e}$}}_t = (\vvarepsilon_1^{\top}, \ldots, \vvarepsilon_t^{\top})^{\top}$ of length $\tilde{n}_t$.
\end{compactitem}
The random effects and errors are assumed to be independent and obey multivariate normal laws, $\ggamma_t \sim \mathcal{N}( \mathbf{0}_p, \sigma_{\gamma}^2 \mathbf{I}_{pp})$ and $\vvarepsilon_t \sim \mathcal{N}( \mathbf{0}_n, \sigma_{\varepsilon}^2 \mathbf{I}_{nn})$ for all $t$, and with zero covariance across data sets, e.g. $\mbox{Cov} ( \vvarepsilon_{t}, \vvarepsilon_{t'}) = \mathbf{0}_{nn}$ for $t \not= t'$.

The maximum likelihood estimator of the above formulated mixed model's fixed regression parameter $\bbeta$ given the variance parameters $\sigma_{\varepsilon}^2$ and $\sigma_{\delta}^2$ is (cf. \citealp{Bate2004}):
\begin{eqnarray*}
\hat{\bbeta}_t^{\mbox{{\tiny (me)}}} & = & [\mathbb{X}^{\top} (  \xi  \mathbb{Z} \mathbb{Z}^{\top} + \mathbf{I}_{\tilde{n}_t, \tilde{n}_t})^{-1} \mathbb{X}]^{-1} \mathbb{X}^{\top} ( \xi  \mathbb{Z} \mathbb{Z}^{\top} + \mathbf{I}_{ \tilde{n}_t, \tilde{n}_t})^{-1} \mathbb{Y},
\end{eqnarray*}
with its first two moments:
\begin{eqnarray*}
\mathbb{E} [\hat{\bbeta}_t^{\mbox{{\tiny (me)}}}] & = & [\mathbb{X}^{\top} (  \xi  \mathbb{Z} \mathbb{Z}^{\top} + \mathbf{I}_{\tilde{n}_t, \tilde{n}_t})^{-1} \mathbb{X}]^{-1} \mathbb{X}^{\top} ( \xi  \mathbb{Z} \mathbb{Z}^{\top} + \mathbf{I}_{\tilde{n}_t, \tilde{n}_t})^{-1} \mathbb{X}_t \bbeta,
\\
\mbox{Var} [\hat{\bbeta}_t^{\mbox{{\tiny (me)}}}] & = & [\mathbb{X}^{\top} (  \xi  \mathbb{Z} \mathbb{Z}^{\top} + \mathbf{I}_{\tilde{n}_t, \tilde{n}_t})^{-1} \mathbb{X}]^{-1} \mathbb{X}^{\top} ( \xi  \mathbb{Z} \mathbb{Z}^{\top} + \mathbf{I}_{\tilde{n}_t, \tilde{n}_t})^{-1}
\\
& & (\sigma^2_{\varepsilon} \mathbf{I}_{\tilde{n}_t, \tilde{n}_t} + \sigma^{2}_{\gamma} \mathbb{Z} \mathbb{Z}^{\top})  ( \xi  \mathbb{Z} \mathbb{Z}^{\top} + \mathbf{I}_{\tilde{n}_t, \tilde{n}_t})^{-1} \mathbb{X}[\mathbb{X}^{\top} (  \xi  \mathbb{Z} \mathbb{Z}^{\top} + \mathbf{I}_{\tilde{n}_t, \tilde{n}_t})^{-1} \mathbb{X}]^{-1} ,
\end{eqnarray*}
in which $\xi = \sigma_{\varepsilon}^{-2} \sigma_{\gamma}^{2}$.

The updated ridge regression estimator and the maximum likelihood estimator of the mixed model's fixed effect parameter can now be compared with respect to their mean squared error. Theorem \ref{theorem.mixed2ridge1} does so for a sequence of studies with orthonormal design matrices.

\begin{theorem} (Mean squared error of mixed vs. updated estimator) \label{theorem.mixed2ridge1} \\
Adopt assumption \textit{A1.linear}. Let $\mathbf{X}_t$ be orthonormal and $\lambda_t >  \sigma_{\varepsilon} (\sigma_{\varepsilon}^2 + \sigma_{\gamma}^2)^{-1/2} 2^{t/2} T^{1/2}$ for $1 \leq t \leq T$.  Then, when initiated by $\hat{\bbeta}_1(\lambda_1) = \hat{\bbeta}_1^{\mbox{{\tiny (me)}}}$, the updated ridge regression estimator outperforms (in the mean squared error sense), the maximum likelihood estimator of the mixed model's fixed effects parameter as given above. Hence, $MSE[ \hat{\bbeta}_T(\lambda_T)] < MSE[\hat{\bbeta}_T^{\mbox{{\tiny (me)}}}]$.
\end{theorem}

\noindent This theorem states that the updated linear regression estimator may, when initiated appropriately and equipped with the right penalty parameter scheme, outperform in the MSE sense the maximum likelihood estimator of the  mixed model's fixed effects parameter. Acknowledged, Theorem \ref{theorem.mixed2ridge1} assumed an orthonormal design matrix for all studies, which is rather restrictive from a practical perspective. In principle, however, a result as in Theorem \ref{theorem.mixed2ridge1} can be obtained for $\mathbf{X}_t = \mathbf{X}$ with $\mathbf{X}$ of sufficient rank. The proof follows that of Theorem \ref{theorem.mixed2ridge1} but the mathematics is more cumbersome and the result less insightful.

Theorem \ref{theorem.mixed2ridge1} assumes the variance parameters $\sigma_{\varepsilon}^2$ and $\sigma_{\delta}^2$ known. In practice, these parameters too need to be estimated from data. This is done in the simulation study that follows.

The performance of the presented updated ridge regression estimator is compared to the maximum likelihood estimator of the mixed model's fixed effect parameter in two simulations. The first simulation investigates the result of Theorem \ref{theorem.mixed2ridge1} under more realistic assumptions. The second simulation reveals a better performance of the updated ridge regression estimator than the maximum likelihood estimator of the mixed model's fixed effect parameter when for a particular study in the sequence accidently data from a different model are sampled. This accidently erroneous sampling from a different model will introduce bias in the latter estimator, as it cannot reduce the influence of the data from this study. The updated ridge regression estimator, however, decides by the size of the penalty parameter how it weighs the current data against the regression estimate obtained from the preceding studies. In principle, it may favor the latter, effectively ignoring the data from the erroneously sampled study. This may cause a small reduction in efficiency compared to its competitor, but this is expected to be outweighed by the avoided bias. Both simulations are modified from that presented in Section \ref{sect:comparison2regularRidge}. The settings are identical except for the fact that $\sigma_{\varepsilon}^2 = 1$. Moreover, in a second scenario the $t$-th data set with $t \in \{ t \, : \, t \mod 10 = 0 \}$ is sampled from an empty model, i.e. $\mathbf{Y}_t = \vvarepsilon_t$. With the arrival of each new data set the parameter $\bbeta$ is estimated using the ML estimator of the mixed model's fixed parameter (from that $t$ when the accumulated sample size exceeds the dimension). But also using the updated ridge regression estimator initiated uninformatively with $\bbeta_0 = \mathbf{0}_p$ and informatively with $\bbeta_0 = \bbeta$. Its penalty parameter is chosen through constrained cross-validation to prevent the detoriation of the fit on the preceding data sets. The average (over hundred repeats) quadratic loss, e.g. $\| \hat{\bbeta}_t [\lambda_t, \hat{\bbeta}_{t-1} (\lambda_{t-1}, \hat{\bbeta}_{t-2})] - \bbeta \|_2^2$, of the mixed model's fixed parameter ML and updated ridge regression estimators are plotted against $t$ (bottom panels of Figure \ref{fig.regular2updateRidge}). In the first scenario with all data sets drawn from the same linear regression model the maximum likelihood estimator of the mixed model's fixed parameter performs better than the uninformatively initiated updated ridge estimator but worse than the informatively initiated one. For small $t$ this picture is also seen in the second scenario with every tenth data set sampled from an empty model. But the maximum likelihood estimator of the mixed model's fixed parameter suffers most (relatively) from the `outlying' data sets, and is for larger $t$ even overtaken by the uninformatively initiated updated ridge regression estimator.

\subsection{State space model}
A commonly applied form of transfer regression learning is the linear Gaussian state space model with time-varying coefficients \citep{Durb2012}. The model comprises an \textit{observation equation}, describing the observations $\mathbf{Y}_t$ at each $t$ by a linear regression model, and a \textit{state equation}, describing the fluctuations in the regression parameter over time. A state space model that applies to the situation studied here is:
\begin{eqnarray*}
\left\{
\begin{array}{lcl}
\mathbf{Y}_t & = & \mathbf{X}_t \bbeta_t + \vvarepsilon_t,
\\
\bbeta_t & = & \textcolor{white}{\mathbf{X}_t} \bbeta_{\textcolor{white}{t}} + \ddelta_t,
\end{array}
\right.
\end{eqnarray*}
where $\vvarepsilon_t \sim \mathcal{N}(\mathbf{0}_{n}, \sigma^2_{\varepsilon} \mathbf{I}_{nn})$ and $\ddelta_t \sim \mathcal{N}(\mathbf{0}_{p}, \sigma^2_{\delta} \mathbf{I}_{pp})$ both identically and independently distributed over $t=1, \ldots, T$ as well as between each other. The estimator of $\bbeta$ obtained through likelihood maximization coincides with the mixed model's one. Hence, we refer to the the previous subsection.

\section{Application}
The \textit{fingertips} study registers annual information on a range of public health indicators for the counties of England (\url{https://fingertips.phe.org.
uk/}). Following \cite{Schaik2019}, who advocate the use of the \textit{fingertips} data for the illustration of novel statistical techniques, we apply the proposed transfer learning procedure to relate the counties' suicide rate to the other health indicators. Building on the \texttt{R}-script of \cite{Schaik2019} the data are downloaded using the \texttt{fingertipsR}-package \citep{Fox2019}. Preprocessing of the data amounts to removal of cases without full information on the health indicators registered. This is done per year as the set of registered health indicators varies over the years. In particular, the size of this set tends to increase over time. Moreover, the data are zero-centered year-wise. The resulting data comprise the years 2008 to 2016, over these years the number of health indicators ranges from 1 to 23 from a total of 23 different while the number of counties with full case information ranges from 57 to 159 over the years (refer to the Supplementary Material III for more details). 

\begin{figure*}[h!]
\centering
\begin{tabular}{ll}
\includegraphics[angle=0, scale=0.41]{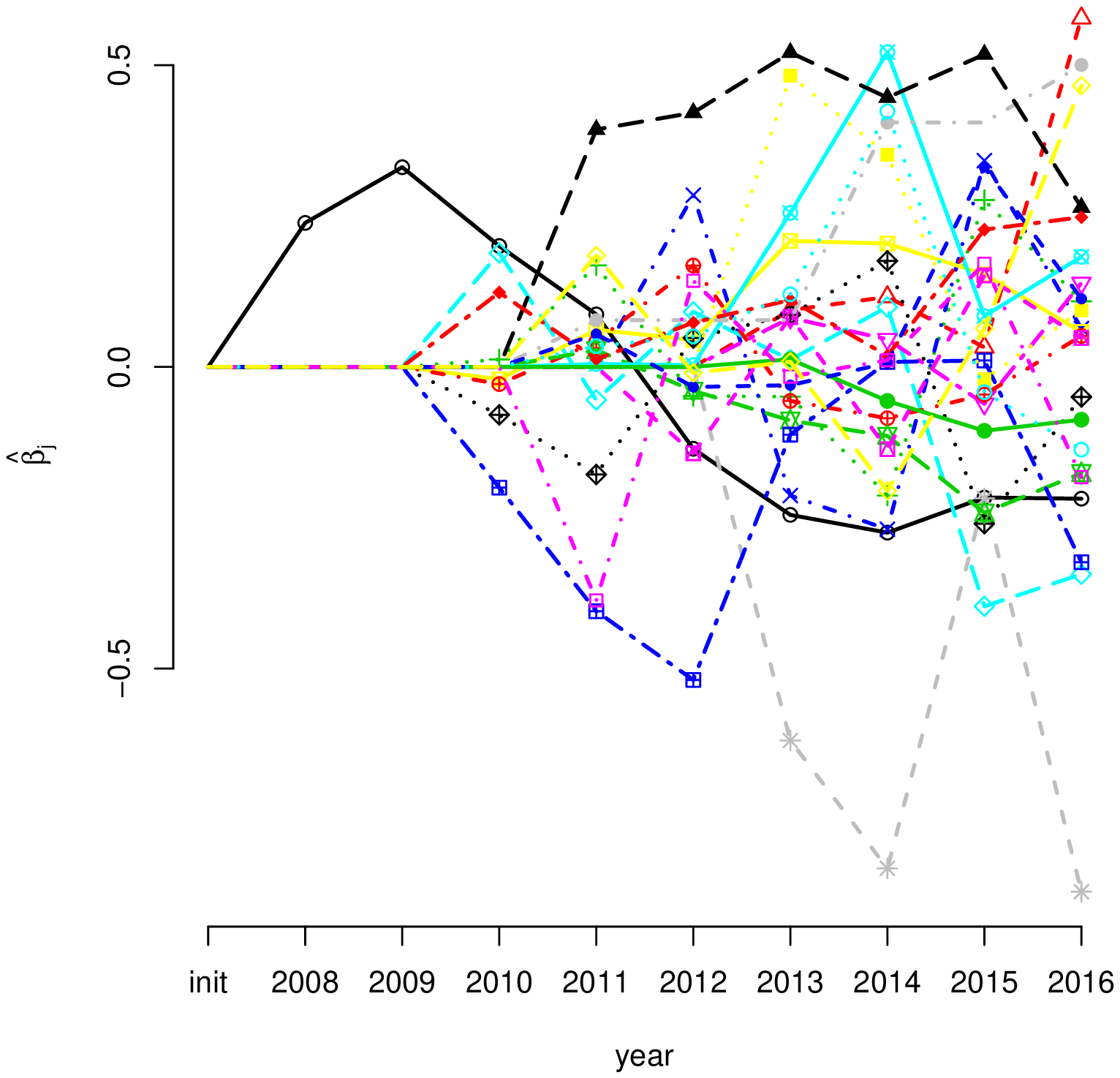}
&
\includegraphics[angle=0, scale=0.41]{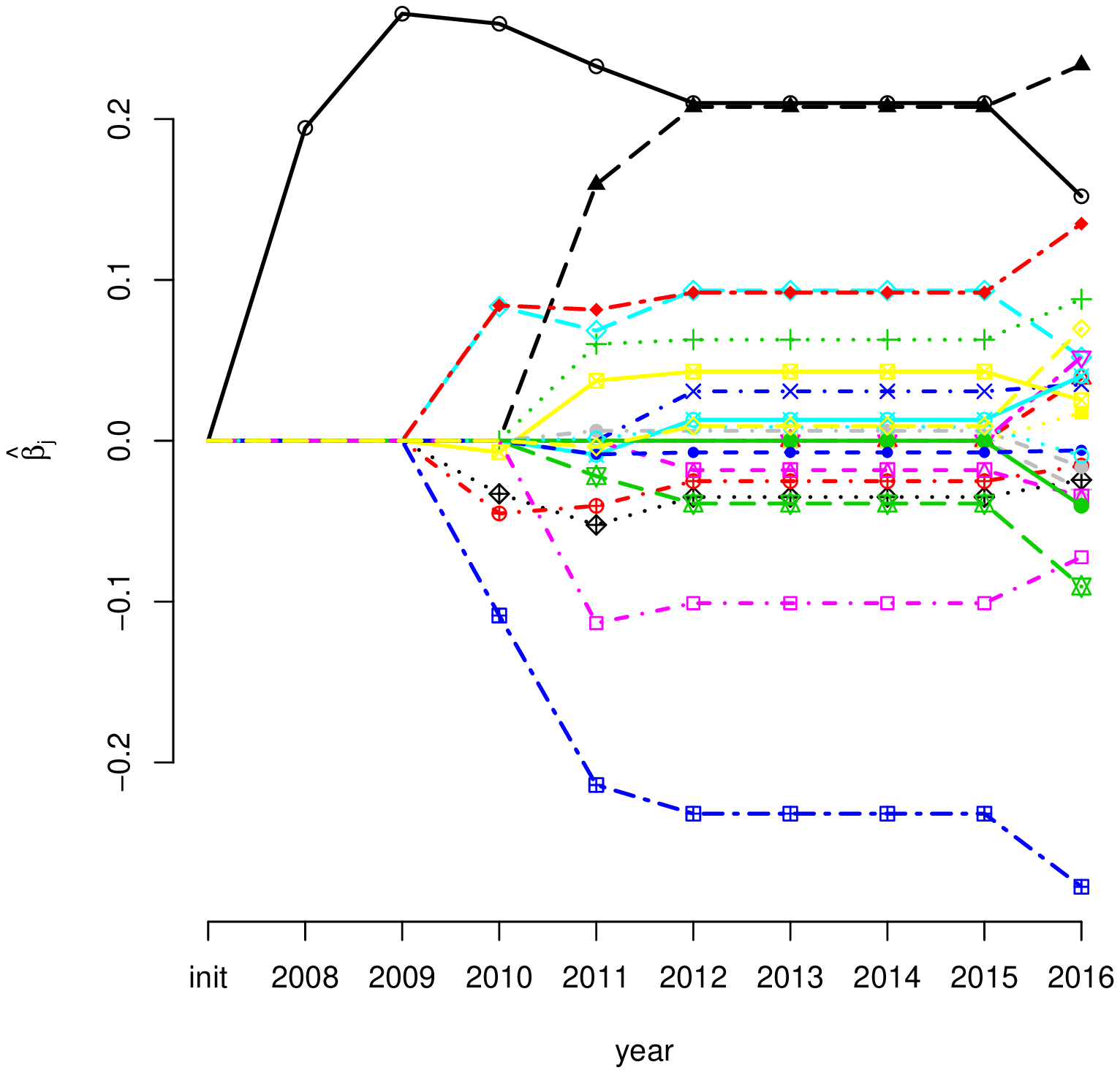}
\end{tabular}
\caption{The panels show the trajectories of the ML (left) and the updated ridge regression, with its penalty parameter chosen via constrained LOOCV, estimates. Each trajectory represents a single covariate. The presence of a health indicator in the data of a particular year is evident from a symbol on its trajectory at the corresponding year. The symbol is omitted in years that the health indicator was not registered.} \label{fig.fingertipsResults}
\hfill{}
\end{figure*}

Analysis of the data commences with the first year available, i.e. 2008. The linear model is fitted with the targeted ridge regression estimator with a zero target, i.e. $\bbeta_0 = \mathbf{0}_p$, and the penalty parameter is chosen by LOOCV. For subsequent years the same estimator is used but the target is formed element-wise from previous estimates: the $j$-th element of the target is taken from the most recent estimate. If the $j$-th covariate was not present in the data preceeding year it is taken from the year before that, and so on, until it is taken from the initial target for the year 2008. Moreover, the penalty parameter is using by unrestricted and constrained LOOCV (as discussed in Section \ref{sect.CV}). Finally, for the purpose of reference the regression parameter is also estimated year-wise with the maximum likelihood regression estimator.

The trajectories of the resulting estimates are shown in Figure \ref{fig.fingertipsResults} and the Supplementary Material III. The main takeaway of the plots is the volatile behaviour of the ML regression estimates. In part this is due to a different sample size and, probably more important, the varying set of health indicators registered each year. The updated ridge regression estimator yield, irrespectively of the employed cross-validation method, yield much smoother trajectories. Hence, providing last year's parameter estimate as a suggestion for the estimation of the current year's one appears to harness against the involvement of a different set of covariates. Indirectly, the usefulness of this suggestion for the learning of the updating of the estimate is also evident by the increase of both the un- and constrained cross-validated penalty parameters over the years (not shown). This indicates that the suggestion is becoming more relevant as years go by. A convenient consequence of the updated ridge regression estimator's smooth trajectories is the consistency of the sign of the regression parameter's estimates over the years, most obvious from the estimators' trajectories for two elements of the regression parameter (provided in the Supplementary Material III), facilitating a sensible interpretation.  Of course, there is no free lunch. The updated ridge rergession estimators exhibit a small detoriation in the fit as can be witnessed from the residuals (see Supplementary Material III), although this appears to be negligible.

\section{Conclusion}
We presented methodology for continuously learning regression models from studies executed over large periods of time with data coming in bit-by-bit as they run their course. It could be considered a frequentist analogy to Bayesian updating and comprises sequential targeted ridge penalized estimation, shrinking towards the latest estimate, when novel data become available. At each update the penalty parameter represents the extent to which the latest estimate yields an adequate model of the novel data. The iteratively updated estimator and its associated linear predictor have been shown to be asymptotically unbiased and consistent. The penalty parameter is chosen through cross-validation in which the search domain of the penalty parameter is constrained to ensure that the newly learned estimate causes little to no detoriation of the fit of the historic data in comparison to the previous estimate. In a comparison with other standard statistical transfer learning procedures situations were identified where these procedures where outperformed by the proposed method. Finally, the proposed method was illustrated on data from an epidemiology study, showing its promise. 

The current transfer learning proposal employs a single penalty parameter per update. Hence, all elements of the novel estimate are shrunken in the same amount to the current one. However, it may be that certain elements of the current estimate provide a better fit than others on the novel data. Then, to differentiate the shrinkage among the novel estimate's elements the single-parameter penalty is to be replaced by a multi-parameter one: $[\bbeta - \hat{\bbeta}_{t-1}(\LLambda_{t-1}, \hat{\bbeta}_{t-1} )]^{\top} \LLambda_{t} [\bbeta - \hat{\bbeta}_{t-1}(\LLambda_{t-1}, \hat{\bbeta}_{t-1} )]$, the penalty parameter $\LLambda_{t}$ is now a matrix.  This matrix may be unstructured, leaving one with the problem of selecting $\tfrac{1}{2} p (p+1)$ penalty parameters. More practically, it may be diagonal and possibly parameterized by a few scalar penalty parameters, thereby shrinking groups of elements similarly. Alternatively, a parametrization in line with generalized ridge regression \citep{Hoer1970, Hemm1975, Lawl1981}, only shrinking differentially in the $p$ canonical directions, may be considered. Either choice provides greater flexibility in the information propogated between updates.

\bibliographystyle{Chigaco}
\bibliography{/home/wessel/Research/Articles/genomics_literature}

\end{document}